\documentclass[10pt,letterpaper]{article}
\usepackage{amsmath}
\usepackage{amssymb}
\usepackage{color}
\usepackage{xr-hyper}
\ifx\pdfoutput\undefined

\usepackage
[dvips,
 verbose=false,
 bookmarks=true,
 colorlinks=true,
 linkcolor=webred,
 filecolor=webbrown,
 citecolor=webgreen,
 pagecolor=webblue,
 urlcolor=webblue,
 pdftitle={Lattice polarized toric K3 surfaces},
 pdfauthor={Falk Rohsiepe},
 pdfsubject={toric geometry, algebraic geometry, string theory,
 mirror symmetry, CFT},
 pdfkeywords={toric algebraic geometry string mirror symmetry CFT},
 bookmarksopen=false,
 pdfpagemode=None,
 pdfview=FitH,
 pdfstartview=FitH,
 extension=pdf]{hyperref}

\else

\usepackage
[pdftex,
 verbose=false,
 bookmarks=true,
 colorlinks=true,
 linkcolor=webred,
 filecolor=webbrown,
 citecolor=webgreen,
 pagecolor=webblue,
 urlcolor=webblue,
 pdftitle={Lattice polarized toric K3 surfaces},
 pdfauthor={Falk Rohsiepe},
 pdfsubject={toric geometry, algebraic geometry, string theory,
 mirror symmetry, CFT},
 pdfkeywords={toric algebraic geometry string mirror symmetry CFT},
 bookmarksopen=false,
 pdfpagemode=None,
 pdfview=FitH,
 pdfstartview=FitH,
 extension=pdf]{hyperref}
\fi

\definecolor{webred}{rgb}{.8,0,0}
\definecolor{webbrown}{rgb}{.6,0,0}
\definecolor{webgreen}{rgb}{0,0.5,0}
\definecolor{webdkgreen}{rgb}{0,0.3,0}
\definecolor{webblue}{rgb}{0,0,0.5}

\newtheorem{thm}{Theorem}[section]
\newtheorem{cor}[thm]{Corollary}
\newtheorem{lem}[thm]{Lemma}
\newtheorem{prop}[thm]{Proposition}
\newtheorem{defn}[thm]{Definition}
\newtheorem{rem}[thm]{Remark}
\newtheorem{exm}[thm]{Example}
\DeclareMathAlphabet\mathscr{U}{eus}{m}{n}
\DeclareMathAlphabet\mathup{OT1}{\rmdefault}{m}{n}

\renewcommand{\[}{\begin{equation}}
\renewcommand{\]}{\end{equation}}
\newcommand{\atopof}[2]{\genfrac{}{}{0pt}{}{#1}{#2}}
\newcommand{\PP}{\ensuremath{\mathbb{P}}}

\newcommand{\ZZ}{\ensuremath{\mathbb{Z}}}
\newcommand{\QQ}{\ensuremath{\mathbb{Q}}}
\newcommand{\CC}{\ensuremath{\mathbb{C}}}

\newcommand{\RR}{\ensuremath{\mathbb{R}}}
\newcommand{\SSS}{\ensuremath{\mathbb{S}}}

\newcommand{\TT}{\ensuremath{\mathscr{T}}}
\newcommand{\TTT}{\ensuremath{\mathbb{T}}}

\newcommand{\OOne}{\text{\textup{1\hspace{-0.25em}l}}}
\newcommand{\OO}{\ensuremath{\mathscr{O}}}
\newcommand{\defas}{\ensuremath{:=}}

\newcommand{\tensor}{\otimes}
\newcommand{\cut}{\cap}

\newcommand{\Deltastar}{{\Delta^\star}}

\newcommand{\codim}{\mathrm{codim}}
\newcommand{\skpd}[2]{\left\langle{#1},{#2}\right\rangle}
\newcommand{\rank}{\mathrm{rk}}

\begin{document}
\rightline{\vbox{\hbox{BONN-TH-2004-06} \hbox{September 2004}}}
\bigskip
\centerline{\LARGE Lattice polarized toric K3 surfaces}
\medskip
\centerline{Falk Rohsiepe\footnote{email:
\href{mailto:rohsiepe@th.physik.uni-bonn.de}{rohsiepe@th.physik.uni-bonn.de}}}
\centerline{Physikalisches Institut} \centerline{der Universit\"at
Bonn} \centerline{Nu{\ss}allee 12, D-53115 Bonn}
\centerline{Germany}

\begin{abstract}
When studying mirror symmetry in the context of K3 surfaces, the hyperk\"ahler
structure of K3 makes the notion of exchanging K\"ahler and
complex moduli ambiguous. On the other hand, the metric is not
renormalized due to the higher amount of
supersymmetry of the underlying superconformal field theory.
Thus one can define
a natural mapping from the classical K3 moduli space to the moduli
space of conformal field theories. Apart from the generalization of mirror
constructions for Calabi-Yau threefolds, there is a formulation
of mirror symmetry in terms of
orthogonal lattices and global moduli space arguments. In many cases
both approaches agree perfectly
--- with a long outstanding exception: Batyrev's mirror
construction for K3 hypersurfaces in toric varieties does not fit
into the lattice picture whenever the Picard group of the K3
surface is not generated by the pullbacks of the equivariant
divisors of the ambient toric variety. In this case, not even the
ranks of the corresponding Picard lattices add up as expected. In this paper
the connection is clarified by refining the lattice picture.
We show (by explicit calculation with a computer) mirror symmetry
for all families of toric K3 hypersurfaces corresponding to dual
reflexive polyhedra, including the formerly problematic cases.
\end{abstract}
\section{Introduction}
Superstring theory at small coupling is described by
superconformal field theories on the string world sheet at central
charge $c=15$. The most simple theory of this kind consists of 10
free bosons (each contributing $1$ to the central charge) and 10
fermions (contributing central charge $1/2$) leading to string
theory in flat 10 dimensional Minkowski space. In order to obtain
theories with fewer (visible) space-time dimensions\footnote{We do
not consider the more modern approach of confining (most of the)
fields to lower dimensional branes.}, one considers products of
(10-2D) dimensional flat Minkowski space-time theories with
internal superconformal field theories with central charge $c =
3D$.

Under certain assumptions\footnote{if the conformal field
theories contain the spectral flow operator in their Hilbert
space} one expects all components of the moduli space of such
theories to contain boundaries consisting of supersymmetric sigma
models on Ricci-flat K\"ahler manifolds $X$ of large radius. This
allows the use of classical geometrical methods for studying these
theories. In the (realistic) case of four flat space-time
dimensions $X$ is a Calabi-Yau threefold. This intimate connection
between quantum field theories and classical geometry led to one
of the most striking predictions concerning the classical
geometries --- mirror symmetry: The deformations of the classical geometry fall into
two seemingly unrelated classes, namely the deformations of the
complex structure labelled by the cohomology group $H^{D-1,1}(X)$ and
deformations of the complexified K\"ahler form\footnote{In the
context of type II string theory one always has to include an
antisymmetric $B$-field, which is only defined modulo integer
classes and conveniently combined with the K\"ahler form to form
the so-called complexified K\"ahler form.} labelled by $H^{1,1}(X)$.

From the viewpoint of conformal field theory these deformations
correspond to marginal operators, which also fall into two
classes, namely $h=\bar{h} = \frac{1}{2}$, $q = \bar{q} = 1$ and
$h=\bar{h} = \frac{1}{2}$, $q = -\bar{q} = 1$. $h$ denotes the
conformal dimension of the marginal operator and $q$ its charge
under the $U(1)$ current contained in the superconformal Virasoro
algebra. Although we also have to deal with two kinds of operators,
the difference seems much less fundamental than for their
geometric counterparts. In fact, changing the sign of the
$U(1)$-charges leads to an isomorphic conformal field theory
\cite{greeneplesser}.

If this field theory also has a geometric
interpretation, it should be a sigma model on another Calabi-Yau
threefold $\tilde{X}$ with $H^{D-1,1}(\tilde{X}) = H^{1,1}(X)$ and
vice versa.

This observation led to the remarkable conjecture,
that Calabi-Yau manifolds should come in pairs with the Hodge
groups exchanged (and, of course, isomorphic underlying
superconformal field theories). Though at first counterintuitive
for mathematicians, lots of evidence for this conjecture has been
found and it is nowadays widely accepted.

For large classes of
Calabi-Yau manifolds recipes for constructing mirror partners are
known, e.g. the quotient construction of \cite{greeneplesser}, its
generalization to hypersurfaces in toric varieties corresponding
to reflexive polyhedra \cite{batdualpoly}, generalizations thereof
\cite{bordualnef,batborci,batconifold} and fiberwise T-Duality
\cite{stromingeryauzaslow}.

If one considers compactifications not to 4, but rather 6 (real)
dimensions, the compactification space has to be either a real
compact 4-torus or a K3 surface. Since K3 surfaces are
hyperk\"ahler manifolds (the underlying superconformal field
theories possess a higher amount of
supersymmetry\footnote{$N=(4,4)$ instead of $N=(2,2)$}), the
(local) factorization of the moduli space into complex and
K\"ahler deformations is lost. Hence, the generalization of the
above mirror symmetry to this case is not so obvious.

On the other
hand, the higher supersymmetry implies nonrenormalization of the
metric, which allows comparison of geometry and conformal field
theory not only near a large radius limit. For the K3 case, the
most beautiful approach to mirror symmetry is given by global
moduli space arguments in terms of orthogonal Picard lattices
\cite{aspinmorrk3mirr,vafamirr,dolgachev}. In fact, this
description is a quantum geometrical version of a much older
duality discovered by Dolgachev, Nikulin and (independently)
Pinkham \cite{dolganicu, pinkham} and used to explain Arnold's
strange duality \cite{arnold}. I will give a short exposition of
this picture in section \ref{orthlattices}.

As already remarked in \cite{dolgachev}, this version of mirror
symmetry fits in nicely with known mirror constructions. A
subtlety arises in the case of Batyrev's mirror construction for
toric hypersurfaces. The mirror symmetry picture drawn in
\cite{aspinmorrk3mirr,dolgachev} fails in the case of nonvanishing
toric correction term (the difference between the ranks of the
Picard groups of the generic K3 hypersurface and the ambient toric
variety). As these toric hypersurfaces are not generic members of
the families corresponding to their Picard lattices, this does not
imply failure of either picture. Nevertheless the failure to match
both pictures in these cases is unsatisfactory. In \cite[Conj.
8.6]{dolgachev} it was conjectured, that in order to obtain a
matching the Picard lattice has to be replaced by a suitably
chosen sublattice. The main purpose of this paper is to explicitly
state the involved lattices and to prove (by computer) the mirror
assertion for all reflexive polyhedra of dimension 3 --- including
the case of nonvanishing correction term.

The layout of the paper is as follows: In section
\ref{orthlattices} I will give a brief exposition of K3 mirror
symmetry in terms of orthogonal lattices, including the necessary
refinements to treat the case of toric hypersurfaces with
nonvanishing toric correction. Section \ref{toricalgo} is devoted
to the calculation of the Picard lattice for toric K3
hypersurfaces. In section \ref{mainandalgo} I will state the main
theorem concerning mirror symmetric families of toric K3
hypersurfaces and describe the algorithm used to prove it.

\section{Moduli spaces and orthogonal lattices}
\label{orthlattices}

\subsection{Moduli space of superconformal field theories and
geometric interpretations}

The exposition in this section roughly follows \cite{nahmlondon}
which is based on
\cite{katrinphd,k3hitch,aspinmorrk3mirr,aspintasi}.

We first recall some well known facts about the space of
Ricci-flat K\"ahler metrics\footnote{As remarked before, for the
description of the moduli space of K3 surfaces the distinction
between complex structure and K\"ahler moduli does not make much
sense.} on $X=K3$. Given such a metric (and an orientation on $X$)
consider the action of the
Hodge star operator $\star$ on the total cohomology. We have $\dim
H^0(X, \RR) = \dim H^4(X, \RR) = 1$ while the odd cohomology
groups vanish. The action of $\star$ on $H^2(X, \RR)$ does not
depend on the scale of the metric but only on the conformal
structure. The wedge product of two elements of $H^2(X, \RR)$
together with the identification $H^4(X, \RR) \cong \RR$ given by
the standard generator yields a scalar product\footnote{This is the same
as the cup product or the intersection product on homology via Poincar\'e
duality and the de Rham isomorphism.} with signature
$(3,19)$. Because of $\star^2 = 1$ on $H^2(X, \RR)$ it splits into
eigenspaces $H^+ \oplus H^-$ with eigenvalues $+1$ and $-1$,
respectively. $H^+$ has dimension 3 and is positive definite,
while $H^-$ has dimension 19 and is negative definite.
The orientation on $X$ induces an orientation on $H^+$.
Since one
can show that Ricci flat metrics with fixed volume are locally
uniquely specified by $H^+ \subset H^2(X,\RR)$,
the tangent space to the corresponding moduli space is given by
$so(H^2(X, \RR)) / (so(H^+) \oplus so(H^-))$. The
Teichm\"uller space of metrics with fixed volume, including
orbifold limits, is the Grassmannian
$$\TT^{3,19} = O^+(3,19) / (SO(3) \times O(19))$$
of oriented positive definite 3-planes $\Sigma\subset H^2(X,
\RR)$. The right hand side gives the connection to the tangent space
description above, where $O^+(3,19)$ is the index 2 subgroup of
$O(3,19)$ containing $SO(3) \times O(19)$.

Now let $\Gamma(3,19)$ be the intersection of $O^+(3,19)$ with the
automorphism group of\footnote{For a K3 surface $H^2(X, \ZZ)$
contains no torsion and therefore $H^2(X, \ZZ) \subset H^2(X,
\RR)$.}  $H^2(X, \ZZ) \subset H^2(X, \RR)$. As $\Gamma(3,19)$ is
the diffeomorphism group of the K3 surface \cite{k3diffeo1,
k3diffeo2, k3diffeo3}, the moduli space of metrics (again
including orbifold limits) is
$$\Gamma(3,19)\backslash \TT^{3,19} \times \RR^+ = \Gamma(3,19)
\backslash O^+(3,19) / (SO(3) \times O(19)) \times \RR^+,$$
 where the extra $\RR^+$ parametrizes the volume.

Turning to the space of superconformal field theories, it has long
been conjectured \cite{seibergconj}, that the duality group
$\Gamma(4,20)$ can be interpreted as automorphism group of the
total integer cohomology $H^\star(X,\ZZ) = H^0(X, \ZZ) \oplus
H^2(X, \ZZ) \oplus H^4(X,\ZZ)$. Correspondingly, the moduli space
of superconformal field theories should be
$$\Gamma(4,20)\backslash \TT^{4,20} = \Gamma(4,20)
\backslash O^+(4,20) / (SO(4) \times O(20)),$$
 where $O^+(4,20)$ should be interpreted as the orthogonal group of
the total cohomology $H^0(X, \RR) \oplus H^2(X, \RR) \oplus
H^4(X,\RR)$. Apart from $H^0(X, \RR) \oplus H^4(X, \RR)$, all
direct sums will be orthogonal.

In \cite{aspinmorrk3mirr} Aspinwall and Morrison were able to
construct the sigma model\footnote{It is more natural to define
the inverse mapping, i.e. to construct the fourplane for
a given sigma model. Our approach just makes invertibility of the
mapping more obvious.\par
One should note, that the existence of all sigma models in the
image of the mapping has not been proven rigorously. However, the
conformal dimensions and operator product expansion coefficients
have a well behaved perturbation expansion in terms of inverse
powers of the volume. We thus make the assumption that a rigorous
treatment is possible.}
 (i.e. metric and $B$-field) for given
positive definite 4-plane $\Xi \in \TT^{4,20}$:

Let us fix an isomorphism $\ZZ(4,20) \cong H^0(X, \ZZ) \oplus
H^2(X, \ZZ) \oplus H^4(X,\ZZ)$ and let $v, v^0$ denote the
standard generators of $H^4(X,\ZZ)$ and $H^0(X,\ZZ)$,
respectively. Furthermore let $\Upsilon = \Xi \cap (H^2(X, \RR)
\oplus H^4(X,\RR))$. Obviously, $\Upsilon$ is a positive definite
3-plane in $H^2(X, \RR) \oplus H^4(X,\RR)$. Its projection to
$H^2(X, \RR)$ yields the wanted positive definite 3-plane
$\Sigma$. $\Sigma$ is three dimensional because $(v)^2 = 0
\Rightarrow v \not\in \Upsilon$ and positive definite because $v
\perp H^2(X, \RR) \oplus H^4(X,\RR)$. Inverting this projection,
one can write
$$\Upsilon = \{\sigma - (B \sigma) v, \sigma \in \Sigma\},$$
 where $B \in H^2(X,\RR)$ (the scalar product on $H^2(X,\ZZ)$ is
nondegenerate) and will be identified with the $B$-field. As yet,
$B$ is only specified up to elements of $\Sigma^\perp$.

If we write $\Xi = \Upsilon \oplus \RR \xi$ and demand $\xi v =
1$, this uniquely determines $\xi$. Using the remaining freedom
for the choice of $B$ it can be written as
$$\xi = v^0 + B + (V-\frac{B^2}{2}) v,\qquad V \in \RR.$$
As $\Xi$ is positive definite, we have $\xi^2 = 2 V > 0$ and $V$
can be identified with the volume. Obviously, the mapping $\Xi
\stackrel{v, v^0}{\mapsto} (\Sigma, B, V)$ is invertible:
$$(\Sigma,B,V) \stackrel{v, v^0}{\mapsto} \langle\xi\rangle_\RR \oplus \underline\xi(\Sigma)$$
where the vector $\xi$ is defined as above and
$$\underline\xi: \left\{\begin{array}{rcl}
H^2(X,\RR) & \rightarrow & H^2(X, \RR) \oplus H^4(X,\RR) \\
\sigma & \mapsto & \sigma - (B \sigma) v
 \end{array}\right..$$

The map $\TT^{4,20} \rightarrow \TT^{3,19}$ given by the
construction indeed isomorphically maps the subgroup of
$\Gamma(4,20)$ leaving both $v$ and $v^0$ invariant onto the
classical automorphism group $\Gamma(3,19)$.

Elements of $\Gamma(4,20)$ leaving $v$ invariant are symmetries of
the superconformal field theory since they correspond to
translations of $B$ by an element of $H^2(X,\ZZ)$.

In order to prove that the whole of $\Gamma(4,20)$ yields
isomorphic superconformal field theories, one has to prove
equivalence for one additional generator, for which two choices
have been made in the literature: T-duality by Nahm und Wendland
\cite{k3hitch} and mirror symmetry\footnote{The complete proof of
this variant has never been published.} by Aspinwall und Morrison
\cite{aspinmorrk3mirr}.

We see that a given sigma model $(\Sigma, B, V)$
specifies a conformal field theory, but the opposite direction
depends on the choice of our sublattice\footnote{together with the labelling
of its isotropic subspaces as $H^0(X,\ZZ)$ and $H^4(X,\ZZ)$. The hyperbolic lattice
$H$ is defined to be $\ZZ^2$ with quadratic form
$\left(\atopof{0}{1} \atopof{1}{0}\right)$.}
$H^0(X,\ZZ) \oplus H^4(X,\ZZ) \cong H \subset \ZZ(4,20)$,
which is called a \textbf{geometric interpretation}\footnote{It
is also customary to call just the choice of $v$ a geometric
interpretation. As remarked above, this differs from our definition
by still allowing for shifts of $B$ by elements of $H^2(X,\ZZ)$, which
has no physical relevance.}.

\subsection{Mirror symmetry}

We now want to study mirror symmetry in this context. As mirror
symmetry is an $N=(2,2)$ phenomenon which in the case
of Calabi-Yau threefolds
involves splitting the moduli space's tangent space into complex
and K\"ahler deformations, we have to be somewhat more specific
than in the above discussion.

For a given oriented 4-plane $\Xi$ (or the corresponding
oriented 3-plane $\Sigma$), we have to
choose a complex structure, which corresponds
to choosing an oriented 2-plane $\Omega \subset \Sigma$ (defined by the real
and imaginary part of the complex structure). This choice
(together with the volume) also fixes the K\"ahler class
compatible with the hyperk\"ahler structure given by $\Sigma$.

In terms of cohomology the choice of $\Omega$ fixes
$H^{2,0}(X,\CC) \oplus H^{0,2}(X,\CC)$. The orthogonal complement
of $\Omega$ in $H^2(X,\RR)$ then yields $H^{1,1}(X,\RR)$ and any
vector $\omega \in H^{1,1}(X,\RR)$ of positive length $2V$
defines a K\"ahler class compatible with the complex structure
given by $\Omega$ and the hyperk\"ahler structure defined
by the oriented 3-plane spanned by $\Omega$ and $\omega$.

Equivalent to the choice of a 2-plane
$\Omega \subset \Sigma$ is the choice of its lift
$\tilde\Omega \subset \Upsilon \subset \Xi$. The latter corresponds to
choosing a specific $N=(2,2)$-subalgebra within the $N=(4,4)$
superconformal algebra. More specifically, it corresponds to
choosing a Cartan torus $u(1)_l \oplus u(1)_r$ of
$su(2)_l \oplus su(2)_r$, where the rotations of $\Xi$ in the twoplane
$\tilde{\Omega}$ are generated by $u(1)_{l+r}$ and those in the
orthogonal plane $\mho \defas \tilde{\Omega}^\perp \subset \Xi$ are
generated by $u(1)_{l-r}$ ($SU(2)_l \times SU(2)_r$ operates on
$\Xi$ via the covering map $SU(2)_l \times SU(2)_r \rightarrow SO(4)$,
for details c.f.\ \cite{k3hitch, katrinphd}).

The choice of $\tilde{\Omega}$ also induces a local splitting of the
moduli space into deformations of $\mho \subset {\tilde\Omega}^\perp
\subset H^\star(X,\RR)$ and deformations of ${\tilde\Omega} \subset
\mho^\perp \subset H^\star(X,\RR)$.

We now want to specify what we mean by a K3 mirror symmetry. On the
level of $N=(2,2)$ superconformal quantum field theories we know what we
want to call a mirror symmetry, namely $u(1)_r \leftrightarrow -u(1)_r$ or
equivalently $u(1)_{l+r} \leftrightarrow u(1)_{l-r}$. In our description,
this corresponds to the interchange $\tilde\Omega \leftrightarrow \mho$ and
leads to the following definition:

\begin{defn}
Let $(\Xi = \tilde\Omega \oplus \mho, \ZZ v \oplus \ZZ v^0)$
be the superconformal field theory corresponding to $(\Sigma, B, V)$ under
the geometric interpretation given by $v, v^0$, together with a choice of an
$N=(2,2)$-subalgebra given by $\tilde{\Omega}$. Further let
$(\Xi' = \tilde\Omega' \oplus \mho', \ZZ v \oplus \ZZ v^0)$ be the theory
corresponding to $(\Sigma', B', V')$ with the choice of
the $N=(2,2)$-subalgebra given by $\tilde{\Omega}'$. If $\gamma(\tilde{\Omega}) = \mho'$
(and vice versa) for an involution $\gamma \in \Gamma(4,20)$, the pair
$\left( (\Sigma, B, V, \tilde\Omega), (\Sigma', B', V', \tilde\Omega') \right)$
is called a \textbf{quantum mirror pair} and
$\gamma$ a \textbf{quantum mirror symmetry}.
\end{defn}

\begin{rem} Apart from classical symmetries, the above transformation
can be interpreted as looking at the old fourplane $\tilde\Omega
\oplus \mho$ from the viewpoint\footnote{The difference is
analogous to the difference between rotating an object or the
coordinate system used to measure its position.} of a new
geometric interpretation $\ZZ \gamma v \oplus \ZZ \gamma v^0$.
\end{rem}

Obviously, this definition is not very restrictive. In particular, the trivial
symmetry $\OOne \in \Gamma(4,20)$ is a quantum mirror symmetry.

This is caused by the fact that we almost completely ignored geometry:
Over any point in the moduli space of $N=(4,4)$ theories one has an
$\SSS^2 \times \SSS^2$ of $N = (2,2)$ theories corresponding to different
choices of the $N=(2,2)$ subalgebra. In the geometrical context, such a choice is
induced by the choice of complex structure, for which one only has one $\SSS^2$
to choose from. We thus arrive at the following definition:

\begin{defn}
A quantum mirror pair $\left( (\Sigma, B, V, \tilde\Omega),
(\Sigma', B', V', \tilde\Omega') \right)$
is called a \textbf{geometric mirror pair} if both
$\tilde\Omega$ and $\tilde\Omega'$ are preimages of complex structures
$\Omega \subset \Sigma$ and $\Omega' \subset \Sigma'$.
In this case, $\gamma$ is called a \textbf{geometric mirror symmetry}.
\end{defn}

\begin{prop} \label{geombyxi}
Let $\Xi$ be a fourplane as above, $\ZZ v \oplus \ZZ v^0$ a
geometric interpretation and $\gamma \in \Gamma(4,20)$
an involution. Let $\xi \in \Xi$ be the vector defined by
$\ZZ v \oplus \ZZ v^0$ as above and $\xi' \in \Xi$ the vector defined
by the dual geometric interpretation
$\ZZ \gamma v \oplus \ZZ \gamma v^0$. Then $\gamma$ gives rise to
a geometric mirror symmetry with respect to $\Xi$ and
$\ZZ v \oplus \ZZ v^0$, iff $\xi\,\perp\,\xi'$.
\end{prop}
\textbf{Proof:} It is clear, that for a geometric mirror pair,
$\xi \in \mho$ and $\xi' \in \tilde\Omega$. For the other
direction, choose $\xi^0 \in \langle \xi,\xi'\rangle^\perp \subset \Xi$
and define $\mho \defas \langle \xi, \xi^0 \rangle$.\hfill $\Box$
\begin{rem}
Proposition \ref{geombyxi} implies, that neither the trivial symmetry
nor the T-duality $v \leftrightarrow v^0$ are geometric mirror
symmetries.
\end{rem}
If we want $\gamma$ to still give rise to a geometric mirror symmetry
when changing only the volume on either side of a given geometric mirror
pair\footnote{i.e.\ we want to obtain another geometric mirror pair
by changing either volume and adjusting the mirror partner accordingly},
we must have $v \perp \gamma v$:
$$\forall V:\,\xi = v^0+B+(V-\frac{B^2}{2})v \perp \gamma v
\;\Rightarrow v \perp \gamma v.$$
It is useful to also demand $v^0 \perp \gamma v$, which implies
$\langle \gamma v^0, v\rangle = \langle \gamma v^0, \gamma^2 v\rangle =
\langle v^0, \gamma v\rangle = 0$. Obviously, this additional condition
can be fulfilled by an appropriate integer $B$-field shift.
For this situation, we obtain

\begin{cor} \label{thisismirror}
Let $\Xi$, $\ZZ v \oplus \ZZ v^0$ be as above and let
$\ZZ w \oplus \ZZ w^0\,\subset\,H^2(X,\ZZ) \defas (\ZZ v \oplus \ZZ v^0)^\perp
\subset H^\star(X,\ZZ)$ be a primitive hyperbolic sublattice.
Assume $B\,\perp\, w$.
Let $\gamma \in \Gamma(4,20)$ be the involution given by exchanging
$v \leftrightarrow w$ and $v^0 \leftrightarrow w^0$.
Then $\gamma$ gives rise to a geometric mirror symmetry.
\end{cor}
\textbf{Proof:} With $B\,\perp\, w$ we have
$\xi \,\perp\,w$ and thus $\xi \in \Upsilon'$,
where $\Upsilon'$ is given by the geometric interpretation
$\ZZ w \oplus \ZZ w^0$.
$\xi' \perp \Upsilon' \,\Rightarrow\,\xi'\perp\xi$.\hfill $\Box$

\vspace{0.5cm}
Even with the latter definition, we do not have a complete analogy to
mirror symmetry for Calabi-Yau threefolds. Note in particular,
that deformations of $\Xi$ fixing $\tilde{\Omega}$ also fix $\Omega$,
but the reverse is not true. Hence, the local splitting of the
moduli space is not the same as distinguishing between deformations of the
complex structure and deformations of K\"ahler form and $B$-field.

The deformations of $\Xi$ fixing $\tilde\Omega$ are given
by the deformation of $V$, the 20 deformations of $B$ perpendicular to
$\Omega$ and the 19 deformations of $\Sigma$ fixing $\Omega$. This
almost looks like complexified K\"ahler deformations, the analogy
being perfect when $B$ is a $(1,1)$-form, i.e. $B \in \Omega^\perp$.

On the other hand,
the deformations fixing $\mho$ contain the 38 deformations of $\Omega$
fixing its orthogonal complement in $\Sigma$, but also contain two
deformations of $\tilde\Omega$ in the direction of
$v^0 + B - (V+\frac{B^2}{2}) v$, which render it useless as preimage of
a complex structure.

In order to preserve such an interpretation, the
corresponding deformation of $\Xi$ has to be accompanied by a deformation
of $\mho$. This effectively means replacing these two deformations
by $B$-field shifts in the direction of $\Omega$.

As remarked above, the analogy to the threefold case can be
enhanced by demanding $B \in \Omega^\perp$. For any $\Xi$, this
can be achieved by choosing $\Omega = \tilde\Omega \subset H^2(X,
\RR)$. For $B\not=0$ this uniquely determines $\tilde\Omega =
\Upsilon \cut H^2(X, \RR) = \Xi \cut H^2(X, \RR)$.

Such values of the $B$-field naturally arise in the context of
algebraic K3 surfaces when restricting complexified K\"ahler forms
on the ambient space to the K3 surface. By abuse of language,
we will hence call such values of the $B$-field \textbf{algebraic}.

\begin{rem}
In this context, a sublety
arises: It is always possible to choose $\Omega$ in such a way
that the K3 surface becomes algebraic. To this end, one simply has
to choose $\Omega$ to be orthogonal to any vector $\rho \in H^2(X,\ZZ)$
with $\rho^2 > 0$ \cite{kodairaalg} leading to a countable\footnote{The
set of choices is countable since the orthogonal complement of $\Sigma$ in
$H^2(X,\ZZ)$ is negative definite.} infinity of
such choices. In general though, none of these choices is compatible
with the choice leading to an algebraic $B$-field.

There has been some confusion in the literature regarding the notion of
a nonalgebraic deformation. The above discussion shows that it \emph{does}
make sense when imposing both conditions.
\end{rem}

\begin{defn}
A geometric mirror pair $\left( (\Sigma, B, V, \tilde\Omega),
(\Sigma', B', V', \tilde\Omega') \right)$
is called an \textbf{algebraic mirror pair} if
$B$ and $B'$ are algebraic $B$-fields (i.e. $\Omega = \tilde\Omega$
and $\Omega' = \tilde\Omega'$) and both $(\Sigma, V, \Omega)$ and
$(\Sigma', V', \Omega')$ are algebraic K3 surfaces.
In this case, the geometric mirror symmetry $\gamma$ is called
an \textbf{algebraic mirror symmetry}.
\end{defn}

The full analogy to the threefold case can be obtained by restricting
to families of K3 sigma models where $\mho$ remains inside $(\gamma v)^\perp$,
which means $B \perp \gamma v$ for
$(\ZZ v \oplus \ZZ v^0) \perp (\ZZ \gamma v \oplus \ZZ \gamma v^0)$.

\begin{defn}
Two families of K3 sigma models with chosen complex structures
are called (quantum, geometric, algebraic)
\textbf{mirror families}, if they are mapped to each other by a fixed
(quantum, geometric, algebraic) mirror symmetry.
\end{defn}

\subsection{Lattice polarized surfaces and orthogonal Picard lattices}
\label{mirsymorthpic}
 We now consider K3 surfaces $X$ with complex structure
$\Omega$, K\"ahler form $\omega$ and algebraic $B$-field $B$,
i.e.\ $\Omega = \tilde\Omega$.

As $H^2(X,\ZZ)$ is an even selfdual lattice with signature
$(3,19)$, it is uniquely determined to be
\[H^2(X,\ZZ) = -E_8 \oplus -E_8 \oplus H \oplus H \oplus H,
\label{latticesplit}\] where the scalar product of the lattice
$-E_8$ is given by minus the Cartan matrix of $E_8$ and $H$ is the
hyperbolic lattice. Now define
$\tilde{H}^2(X,\ZZ)$ by $\tilde{H}^2(X,\ZZ) \oplus \tilde{H}
\defas H^2(X,\ZZ)$,
where $\tilde{H}$ is one of the hyperbolic lattices from
(\ref{latticesplit}). For later convenience, we define $\hat{H}
\defas H^0(X,\ZZ) \oplus H^4(X,\ZZ)$.

By the Lefschetz theorem the Picard lattice of a K3 surface is
given by $Pic(X) = H^{1,1}(X,\RR) \cut H^2(X,\ZZ)$, the Picard
number (its rank) is denoted by $\rho(X)$. Using the notations of
the preceding section we therefore have
$$Pic(X) = H^2(X,\ZZ) \cut \Omega^\perp.$$
We now consider primitive nondegenerate sublattices $M_1 \subseteq
M_2 \subset \tilde{H}^2(X,\ZZ)$ with ranks $\rho_1$ and\ $\rho_2$,
respectively.

For given sublattices $M_1,M_2$ we consider the family
$K3_{M_1,M_2}$ of all sigma models on $X$ with complex structure
$\Omega$, K\"ahler form $\omega$ and B-field $B$, subject to the
following conditions\footnote{By $M_{1/2}^\perp \subset X$ we
denote the orthogonal complement inside $X$, while $M_{1/2}^\perp$
alone always means $M_{1/2}^\perp \subset \tilde{H}^2(X,\ZZ)$.}:
\begin{enumerate}
\item $M_2 \subseteq Pic(X) = \Omega^\perp\cut
H^2(X,\ZZ)$ \hfill $\Leftrightarrow\quad\quad \Omega \subset
(M_2^\perp \oplus \tilde{H}) \otimes \RR$\phantom{.}
\item $B, \omega \in M_1 \otimes \RR$ \hfill
$\Leftrightarrow\quad\quad \mho \subset
(M_1^{\mbox{\phantom{\scriptsize$\perp$}}} \oplus \hat{H}) \otimes
\RR$.
\end{enumerate}

\begin{rem}
Whenever $B \not= 0$, $\Omega$ and $\mho$ are uniquely specified
by the fourplane $\Xi$ and the condition $\Omega \subseteq \Xi \cap
H^2(X,\RR)$. Otherwise, the two conditions select a complex
structure in $\Xi$.
\end{rem}
\begin{rem}
If the family $K3_{M_1,M_2}$ is not empty, $M_1$ must obviously contain
a vector with positive length squared. Hence, $K3_{M_1,M_2}$
consists of algebraic K3 surfaces with algebraic $B$-fields.
\end{rem}
\begin{rem}
In the language of \cite{dolgachev} condition 1 says that we
consider $M_2$-polarized K3 surfaces. Condition 2 poses additional
constraints on the choice of B-field and K\"ahler form.
\end{rem}

\begin{rem} \label{canignore}
Splitting off $\tilde{H}$ is quite a loss of
generality in comparison with the discussion in \cite{dolgachev}.
It guarantees\footnote{In fact, it is equivalent: The existence of a
1-admissible vector by definition implies the existence of a primitive
hyperbolic sublattice of $M_2^\perp \subset H^2(X,\ZZ)$, which is
unique up to isometry of $H^2(X,\ZZ)$ \cite[Proposition
5.1]{dolgachev}.} (in the language of \cite{dolgachev}) the
existence of a 1-admissible isotropic vector in $M_2^\perp \subset
H^2(X,\ZZ)$. It also implies \cite[Proposition 5.6]{dolgachev}
irreducibility of the moduli space of $M_2$-polarized K3 surfaces
and enables us to ignore the orientation of $\Sigma$ for the
following discussions. The more general case of $n$-admissible vectors
in \cite{dolgachev} obviously does not correspond to a mirror
symmetry in our definition (or any definition demanding isomorphic
conformal field theories) as such vectors do not define geometric
interpretations.
\end{rem}

We now consider the algebraic mirror symmetry $\gamma$ given by exchanging $v, v^0$
with the generators of  $\tilde{H}$. The corresponding mirror pairs are
given by
$$\left( (\langle \Omega, \omega\rangle_\RR, B, V, \Omega),
(\Sigma', B', V', \Omega') \right),$$
where $\Sigma',B',V'$ are determined by $\gamma\Xi,v,v^0$
and $\Omega' = \gamma \mho$. $\Omega'$ indeed defines a complex structure since
\begin{eqnarray*}
\Omega & \subseteq & \Xi\;\cap\;\hat{H}^\perp \subset H^\star(X,\ZZ) \otimes \RR\; \:\mbox{and} \\
\mho & \subseteq & \Xi\;\cap\;\tilde{H}^\perp \subset H^\star(X,\ZZ) \otimes \RR.
\end{eqnarray*}

As the projection of $\mho$ to $H^2(X,\RR)$ is spanned by $B$ and
$\omega$,
$$Pic(X') \supseteq M_1^\perp$$
and
$$B',\omega' \in M_2^\perp \otimes \RR.$$
The same argument works in both directions\footnote{We use the
(more or less trivial) fact, that $((M_\RR)^\perp \cap
\tilde{H}^2(X,\ZZ)) \otimes \RR = (M_\RR)^\perp$ for any sublattice
$M\subset \tilde{H}^2(X,\ZZ)$ ($(M_\RR)^\perp$ is the kernel of
a matrix with integer entries).} and we obtain algebraic mirror
symmetry for the families
$$K3_{M_1,M_2} \leftrightarrow K3_{M_2^\perp, M_1^\perp}$$
as a generalization of the mirror symmetry
$$K3_{M_1,M_1} \leftrightarrow K3_{M_1^\perp, M_1^\perp}$$
in \cite{aspinmorrk3mirr}.

\begin{rem}
As long as we restrict ourselves to $M_1 = M_2$, generic members
$X \in K3_{M_1,M_1}, X' \in K3_{M_1^\perp, M_1^\perp}$ of the two
mirror symmetric families fulfill
$$\rho(X) = \dim M_1 = \rho_1 = 20 - \dim M_1^\perp = 20 - \rho(X'),$$
the widely used landmark for mirror symmetric families of K3
surfaces. This formula is obviously not correct, if $M_1 \not=
M_2$.
\end{rem}

\section{Picard lattices for toric K3 hypersurfaces}
\label{toricalgo}

\subsection{Toric preliminaries}

We use (almost) standard notations as follows\footnote{For an
introduction to toric geometry I recommend
\cite{fulton,oda,coxrecent}}. Let $N \cong \ZZ^3$ and $M = N^\ast$
be dual three dimensional lattices, $\Delta \subset M_\RR$ a
reflexive polyhedron as defined in \cite{batdualpoly} and
$\Delta^\ast \subset N_\RR$ its polar dual. Let $X_\Delta$ denote
the toric variety corresponding to some maximal
crepant\footnote{i.e. the resolution of singularities $X_\Delta =
X_\Sigma \rightarrow X_{\mathscr{N}(\Delta)}$ preserves the
canonical class.} refinement $\Sigma$ of the normal fan
$\mathscr{N}(\Delta)$ of $\Delta$, i.e. the fan over some maximal
triangulation of $\partial \Delta^\ast \cap N$ (see remark
\ref{onegeomphase} below concerning the choice of triangulation).
Let $(\CC^\star)^3 \cong \TTT^3 \subset X_\Delta$ denote the three
dimensional algebraic torus corresponding to the 0 dimensional
cone in $\Sigma$. Let $\{D_i, i\in I\}$ denote the set of toric divisors
of $X_\Delta$ corresponding to the rays $\{\rho_i \in \Sigma^{(1)}
\subset \Sigma\}$. For each $\rho_i$ let $n_i \in \rho_i$ be its
primitive generator. Let $\OO_{X_\Delta}(\sum_{i\in I} D_i)$ be
the anticanonical line bundle on $X_\Delta$. The set of zeroes
$Z(\chi)$ of a generic global section $\chi \in
\Gamma(\OO_{X_\Delta}(\sum_{i\in I} D_i), X_\Delta)$ is called a
\textbf{toric K3 hypersurface} in $X_\Delta$. Using the
holomorphic quotient construction of $X_\Delta$ the section $\chi$
can be identified with a homogenous polynomial
\[\chi = \sum_{m\in \Delta \cap M} a_m \prod_{n_i, i\in I}
z_i^{\skpd{m}{n_i}+1}\label{explicitpoly}\]
 in the homogenous coordinate ring\footnote{For a discussion
of this ring as well as the holomorphic quotient construction c.f.
\cite{coxrecent}.} $S_\Sigma$ of $X_\Delta$.

The following facts have been proven in \cite{batdualpoly}:
\begin{enumerate}
\item For generic $\chi$, $Z(\chi)$ is $\Sigma$-regular, i.e. the
intersections $Z(\chi) \cap O_\sigma$ are either empty or smooth
with codimension 1 in $O_\sigma$ for all torus orbits $O_\sigma$
corresponding to cones $\sigma \in \Sigma$.
\item For generic $\chi$ (and, as stated above, maximal
triangulation of $\partial \Delta^\ast$), $Z(\chi)$ is a smooth
two dimensional Calabi-Yau variety, i.e. a K3 surface.
\end{enumerate}

\begin{rem}\label{onegeomphase} As toric divisors corresponding to
points $n_i$ on the (codimension 1) faces of $\Delta^\ast$ have
empty intersection with the generic hypersurface (which was
already proven in \cite{batdualpoly} and can easily be checked
using the formulas derived in section \ref{toricpiccalc} below),
one only has to maximally triangulate faces of $\Delta^\ast$ with
codimension $\ge 2$ to obtain a maximal crepant desingularization
of the hypersurface $Z(\chi) \subset X_{\mathscr{N}(\Delta)}$.
Since these are just the vertices and edges of $\Delta^\ast$, this
maximal triangulation is uniquely determined. This implies that
the (usually numerous) cones of the secondary fan for $X_\Delta$
corresponding to smooth K3 surfaces combine to form a single
geometric phase of the associated linear sigma model.

For technical reasons it is nevertheless often convenient to work
with a complete maximal triangulation of $\partial \Delta^\ast$.
The intersection theory on the surface does not depend on the
specific choice.
\end{rem}

We now want to study the Picard lattice of the generic smooth
toric K3 surface for given reflexive Polyhedron $\Delta$. In
\cite{k3chowformula} the formula
 \[ \rho(Z_\chi) =
l(\Delta^\ast) - \dim \Delta - 1 - \sum_{\mathrm{codim}
\Gamma^\ast = 1} l^\ast(\Gamma^\ast) + \sum_{\codim \Gamma^\ast =
2} l^\ast(\Gamma)l^\ast(\Gamma^\ast),\label{picardformula}\]
 for the Picard number of the generic smooth toric K3 hypersurface
$Z_\chi \subset X_\Delta$ was stated by reference to
\cite{batdualpoly}. In (\ref{picardformula}) $\Gamma^\ast$ denotes
a face of $\Delta^\ast$ of the given codimension, $\Gamma$ the
dual face of $\Delta$, $l(\Delta^\ast)$ is the number of integer
points in $\Delta^\ast$ and $l^\ast(\Gamma)$ is the number of
integer points in the relative interior of $\Gamma$.

This is both true and false. In \cite{batdualpoly} the above
formula was proven for the Hodge number $h^{1,1}(V)$ of a smooth
$\Delta$-regular toric Calabi-Yau hypersurface $V$ with $\dim V
\ge 3$. The case $V=K3$ with $h^{1,1}(K3) = 20$ was explicitly
excluded. Nevertheless, the above formula is correct, but one has
to be very careful about the conditions, under which it is true.

\cite[Theorem 4.4.2]{batdualpoly} only demands the Calabi-Yau
hypersurface to be $\Delta$ regular, which can be assured by using
just a subset of the linear system of global sections of the
anticanonical bundle on $X_\Delta$. In particular, one easily sees
that the linear subsystem spanned by the sections corresponding to
vertices of $\Delta$ suffices to carry through the Bertini type
argument leading to $\Delta$ regularity of the Calabi-Yau
hypersurface defined by the generic member. As we will shortly
see, this does \textbf{not} suffice for validity of
(\ref{picardformula}). Rather, the complete linear system is
needed.

\begin{exm}
Consider the \textbf{Quartic line} of K3 hypersurfaces in $\PP^3$
defined by $\chi = \sum_{i = 1}^4 X_i^4 - 4\lambda \prod _{i =
1}^4 X_i,\;\lambda \in \CC$. An easy calculation shows, that all
of these hypersurfaces except for $\lambda$ a fourth root of unity
(where the hypersurface has 16 node singularities inside the three
dimensional open torus $\TTT^3$) are $\Delta$-regular.

The generic Picard number of a quartic in $\PP^3$ as calculated by
(\ref{picardformula}) is
$$\rho(Z_\chi) = 5 - 3 - 1 - 4 \cdot 0 + 6 \cdot 0 \cdot 3 = 1.$$
Using the algebraic automorphism group $\ZZ_4^2$ and methods
developed in \cite{nikulinlemma}, one can show that the generic
Picard number of the quartic line is 19. This clearly shows that
$\Delta$ regularity alone does not suffice.
\end{exm}

What one \emph{does} get from \cite{batdualpoly} is a lower bound
on the generic Picard number: When specialized to the K3 case, the
exact sequence from the proof of \cite[Theorem 4.4.2]{batdualpoly}
reads
$$0 \rightarrow H^{1,1}_c(Z_\chi \cap \TTT^3) \rightarrow
H^{1,1}_c(Z_\chi) \rightarrow H^2_c(Z_\chi \backslash (Z_\chi \cap
\TTT^3)) \rightarrow H^3_c(Z_\chi \cap \TTT^3) \rightarrow
0.\label{batexseq}$$

Now the irreducible components of $Z_\chi \backslash (Z_\chi \cap
\TTT^3)$ clearly are divisors of $Z_\chi$ and because of
$\rank(H^3_c(Z_\chi \cap \TTT^3)) = \rank(H^5_c(\TTT^3)) = 3$ and
the above sequence the space of relations between them has
dimension 3. The number of these components can be counted just as
in higher dimension and for $\Delta$-regular $\chi$ we obtain
 $$\rho(Z_\chi) \ge \tilde\rho(\Delta) \defas l(\Delta^\ast) - 4 - \sum_{\mathrm{codim}
\Gamma^\ast = 1} l^\ast(\Gamma^\ast) + \sum_{\codim \Gamma^\ast =
2} l^\ast(\Gamma)l^\ast(\Gamma^\ast).$$
 In order to obtain an upper bound on the generic Picard number,
we prove the following easy lemma:

\begin{lem}
The number of real deformations of complex structure fixing the
Picard lattice $Pic \subset H^2(K3,\ZZ) \cong \ZZ^{3, 19}$ of a
given K3 variety $V$ with $\rank Pic = \rho$ is $D = 2(20 -
\rho)$.
\end{lem}
\textbf{Proof:} The deformations are just the deformations of the
oriented positive definite two-plane $\Omega \subset Pic^\perp
\subset \RR^{3,19}$ discussed in the preceding sections by
elements of $H^{1,1}(V,\RR)$, i.e.\ by elements of $\RR^{3,19}$
perpendicular to $\Omega$. With $a\in\{0,1\}$ we
obtain\footnote{$a = 1$ for algebraic $V$}
$$D = \dim so(3-a, 19 -\rho + a) - \dim so(1-a, 19 - \rho + a) - \dim so(2) = 2(20 - \rho).$$
\hfill $\Box$

\begin{cor} \label{maxgeneric}
Let $F$ be a family of K3 surfaces with $k$ continuous real
deformation parameters for the complex structure. Then the Picard
number of a generic member $V \in F$ cannot exceed $20 - \lfloor
k/2 \rfloor$.
\end{cor}

\begin{rem}
The condition \emph{generic} in Corollary \ref{maxgeneric} is
stronger than usual in algebraic geometry. It is here used in the
same way as the generic element of $\RR$ is irrational.
\end{rem}

Now, the first term in (\ref{batexseq}) just represents
deformations of the complex structure by deformations of the
defining polynomial. Since $h^{1,1}(Z_\chi) = 20$, we can use
Lemma \ref{maxgeneric} with $k = 2 \dim H^{1,1}_c(Z_\chi \cap
\TTT^3) = 2 (20 - \tilde\rho(\Delta))$. For the K3 hypersurface
$Z_\chi$ with $\chi$ a generic member of the complete linear
system we thus obtain
$$\tilde\rho(\Delta) \le \rho(Z_\chi) \le 20 - \frac{2}{2}(20 -
\tilde\rho(\Delta)) = \tilde\rho(\Delta).$$

\begin{rem}
When considering families of K3 surfaces allowing all deformations
of the complex structure preserving some original Picard lattice
$Pic(V_0)$, points with enlarged Picard lattice for any $\rank
Pic(V) > \rank Pic(V_0)$ are dense. Dimensional counting shows
that this is the case for families of toric K3 hypersurfaces.
\end{rem}

We now want to calculate not only the rank, but the complete
Picard lattice of a toric K3 hypersurface $Z(\chi)$, which amounts
to calculating the Chow group $A^1(Z(\chi))$ and the intersection
pairing
$$\#: A^1(Z(\chi)) \times A^1(Z(\chi)) \stackrel{\cdot}{\rightarrow}
A^2(Z(\chi)) \cong \ZZ.$$

One part of this Chow group is easily calculated and stems from
the Chow group $A^1(X_\Delta)$ of the toric variety itself. This
Chow group is generated by the toric divisors of $X_\Delta$
subject to the linear relations given by the exact sequence
 \[0 \rightarrow M \stackrel{\alpha}{\rightarrow} \bigoplus_{i\in I}
\ZZ D_i \rightarrow A^1(X_\Delta) \rightarrow 0,\label{exseq}\]
 where $\alpha: M \ni m \mapsto \sum_{i\in I} \langle m, n_i
\rangle D_i$ and $\langle .,.\rangle$ is the natural pairing
between $M$ and $N$. As we have chosen a maximal triangulation of
$\partial \Delta^\ast$, three pairwise different toric divisors
$D_i, D_j, D_k$ yield
$$D_i \cdot D_j \cdot D_k =
\left\{\begin{array}{ll} 1 & \mbox{if $\rho_i, \rho_j, \rho_k$
are contained in a single cone $\sigma \in\Sigma$ and} \\
0 & \mbox{otherwise.}
\end{array}\right.$$
We are not interested in this Chow group itself, but rather in its
image under the homomorphism $i^\star: A^\star(X_\Delta)
\rightarrow A^\star(Z(\chi))$, where $i$ denotes the embedding
$Z(\chi) \rightarrow X_\Delta$. For future use we define
$Pic_{\mathrm{tor}}(\Delta) \defas i^\star A^1(X_\Delta)$ for a
generic section $\chi$.

For the intersections $\#(i^\star D_i, i^\star D_j)$ we obtain
$\#(i^\star D_i, i^\star D_j) = i^\star D_i \cdot i^\star D_j =
i_\star (i^\star D_i \cdot i^\star D_j) = i_\star i^\star (D_i
\cdot D_j) = Z(\chi) \cdot D_i \cdot D_j$.

As already mentioned above, divisors $D_i$ corresponding to points
$n_i$ in the interior of codimension 1 faces of $\Delta^\ast$ are
in the kernel of $i^\star$. This is easily checked using $Z(\chi)
= \sum_i D_i$ in $A^1(X_\Delta)$: Let $n_i$ lie in the interior of
the facet $\Gamma^\ast$ of $\Delta^\ast$ given by the inner face
normal $m\in M$ ($\Gamma^\ast = \{n \in \Delta^\ast | \langle m, n
\rangle = -1\}$). Then
$$D_i \cdot Z(\chi) = D_i \cdot \sum_{j\in I} D_j = D_i \cdot
\sum_{n_j \in \Gamma^\ast} D_j.$$
 Now according to (\ref{exseq}) we have
$$0 = \sum_{j\in I} \langle m, n_j \rangle D_j$$
in $A^1(X_\Delta)$ and therefore
$$D_i \cdot Z(\chi) = D_i \cdot \left(\sum_{n_j \in \Gamma^\ast} D_j + \sum_{j\in I}
\langle m, n_j \rangle D_j \right) = D_i \cdot \sum_{n_j \not\in
\Gamma^\ast} \langle m, n_j \rangle D_j = 0.$$

Hence, we will happily ignore these divisors from now on. The rest
of $A^1(X_\Delta)$ maps injectively to $A^1(Z(\chi))$ and the
intersection matrix can be calculated by variations of the above
theme. Though this has already been done in\footnote{The second
author of \cite{k3chowformula} does not want to be cited in this
context, assumingly because derivation of these formulas is just
an application of standard formulas from the literature. Even the
basics necessary to calculate the full intersection matrix as I
will do in section \ref{completepicard} are already contained in
\cite{batdualpoly}.} \cite{k3chowformula}, for completeness of the
exposition I will repeat the calculation in the following
section\footnote{In particular my formula for the
self-intersection number of a divisor belonging to a vertex of
$\Delta^\star$ can be used in all cases and is thus much better
suited to be used in a computer program than the formulas derived
in \cite{k3chowformula}.}.

\subsection{Calculation of the toric Picard
lattice\texorpdfstring{ $Pic_{\mathrm{tor}}$}{}}
\label{toricpiccalc}

We first turn to the intersection of two different divisors
$i^\star D_1$ and $i^\star D_2$. If either $n_1$ or $n_2$ lies in
the interior of a facet of $\Delta^\ast$, the intersection
obviously vanishes. The same holds for $n_1$ and $n_2$ which lie
on different edges of $\Delta^\ast$. If the different edges are
not borders of a common facet, this is obvious. Let us now assume
$n_1$ and $n_2$ belong to a common cone in $\Sigma$. In this case,
they obviously belong to exactly two common cones. We denote the
corresponding third generators by $n_3$ and $n_4$, which lie on
(not necessarily different) facets of $\Delta^\ast$ given by inner
face normals $m_3$ and $m_4$, i.e.\ $\langle m_3, n_3\rangle = -1$
and $\langle m_4, n_4\rangle = -1$. For the intersection we obtain
\begin{eqnarray}
i^\star D_1 \cdot i^\star D_2 & = & \sum_i D_1\cdot D_i\cdot D_2 \label{neighboringformula} \\
& = & D_1\cdot D_1\cdot D_2 + D_1\cdot D_2\cdot D_2 + D_1\cdot
D_3\cdot D_2 + D_1\cdot D_4\cdot D_2 \nonumber
\end{eqnarray}
Because of (\ref{exseq}) we have
 \[0 = \sum_i \skpd{m_3}{n_i}
D_i \Rightarrow D_1 = -D_2 - D_3 + \skpd{m_3}{n_4}D_4 +
\ldots\label{d1altern},\]
 where $D_1\cdot (\ldots)\cdot D_2 = 0$.
If we insert this into the first term in
(\ref{neighboringformula}) (for one $D_1$ only), we obtain
\begin{eqnarray}
i^\star D_1 \cdot i^\star D_2 & = & D_1\cdot D_2\cdot D_4 +
\skpd{m_3}{n_4} D_1\cdot D_2\cdot D_4 \nonumber \\
& = & (1 + \skpd{m_3}{n_4}) D_1\cdot D_2\cdot D_4 \nonumber \\
& = & \skpd{m_3-m_4}{n_4}\label{twointer}.
\end{eqnarray}
If now $n_1$ and $n_2$ lie on different borders of the same facet,
$m_3 = m_4$ and the intersection vanishes\footnote{This is just
what one expects, since intersections on the K3 surface should not
depend on the triangulation of the facets.}.

For two points $n_1$ and $n_2$ on (not necessarily in the interior
of) the same edge $\theta^\ast$ of $\Delta^\ast$ the intersection
can only be nonzero if the two points are neighboring. In this
case and for any maximal triangulation $\partial \Delta^\ast$ they
share two common cones and $m_3$ and $m_4$ are the inner face
normals of the two facets which intersect in $\theta^\ast$.

Because of $\skpd{m_3-m_4}{n_1} = \skpd{m_3-m_4}{n_2} = 0$ and as
$\{n_1,n_2,n_4\}$ form a basis of $N$, (\ref{twointer}) is just
the integer length of $m_3-m_4$, i.e. the length\footnote{To avoid
confusion, we should note that $l(\theta)$ is \emph{not} the
number of integer points in $\theta$ (in contrast to the notation
in (\ref{picardformula})) but rather $\#(\mbox{integer points in
$\theta$}) = l(\theta)+1 = l^\star(\theta)+2$.} $l(\theta)$ of the
edge $\theta$ of $\Delta$ dual to $\theta^\star$:
\[i^\star D_1 \cdot i^\star D_2 = l(\theta).\label{neighboringformula2}\]

We now turn to the self-intersection number for $i^\star D_1$
where $n_1$ lies in the interior of some edge $\theta^\ast$.
Denote its two neighboring points on $\theta^\ast$ by $n_2$ and
$n_3$. Again let one of the neighboring facets be given by inner
face normal $m_3$. Using (\ref{d1altern}) and
(\ref{neighboringformula2}) we obtain
\begin{eqnarray}
i^\star D_1 \cdot i^\star D_1 & = &
D_1\cdot Z(\chi)\cdot\sum_{i\not=1}\skpd{n_i}{m_3}D_i \nonumber \\
& = & - D_1\cdot Z(\chi)\cdot (D_2 + D_3) + 0 \nonumber \\
& = & -2\,l(\theta).
\end{eqnarray}

For the remaining case of a vertex $n_1$ of $\Delta^\ast$ let one
of the facets containing $n_1$ be given by $m_1$, i.e.
$\skpd{m_1}{n_1} = -1$. We then obtain
\begin{eqnarray}
i^\star D_1 \cdot i^\star D_1 & = &
D_1\cdot Z(\chi)\cdot\sum_{i\not=1}\skpd{m_1}{n_i}D_i \nonumber \\
& = & D_1\cdot Z(\chi)\cdot(\sum_{i=2}^k \skpd{m_1}{n_i}D_i) \nonumber \\
& = & \sum_{j=2}^k \skpd{m_1}{n_j} i^\star D_1 \cdot i^\star D_j.
\label{vertexselfintersectform}
\end{eqnarray}
The $i^\star D_1 \cdot i^\star D_j$ have already been calculated
using (\ref{neighboringformula2}).

\subsection{Calculation of the complete Picard lattice}
\label{completepicard}
 If the \textbf{toric correction term} $\delta = \sum_{\codim \Gamma^\ast =
2} l^\ast(\Gamma)l^\ast(\Gamma^\ast)$ in (\ref{picardformula})
vanishes, we have $Pic(\Delta) \defas Pic(Z(\chi)) =
Pic_{\mathrm{tor}}(Z(\chi))$ for generic $\chi$. Now let us assume
$\delta\not=0$ and let $D_i$ be a toric divisor corresponding to a
point $n_i$ in the interior of an edge $\theta^\ast$, for which
the dual edge $\theta$ also contains interior integer points. Then
the intersection of $D_i$ with the K3 hypersurface splits into
$l(\theta)$ disjoint so-called \textbf{nontoric divisors}. This
well known fact is most easily seen by using the homogenous
polynomial description for $\chi$. When restricting to the divisor
under consideration, this polynomial reduces to a polynomial in
one variable, the zeroes of which determine the (for general
$\chi$) disjoint nontoric divisors: Using the description
(\ref{explicitpoly}) and restricting to $D_i$, i.e. setting $z_i =
0$, yields
\begin{eqnarray*}
\chi|_{z_i = 0} & = & \sum_{m\in \Delta \cap M} a_m \prod_{n_j,
j\in I} z_j^{\skpd{m}{n_j}+1} \\
& = & \sum_{m\in \theta \cap M} a_m \prod_{n_j\not\in
\theta^\ast} z_j^{\skpd{m}{n_j}+1} \\
& = & a_{m_0} \prod_{n_j\not\in \theta^\ast}
z_j^{\skpd{m_0}{n_j}+1} \sum_{k = 0}^{l(\theta)}
\frac{a_{m_k}}{a_{m_0}} r^k
\end{eqnarray*}
where $m_0, \ldots, m_{l(\theta)}$ denote the integer points along
$\theta$ and $r$ is the rational function on $X_\Delta$ defined by
$r = \prod_{j \in I} z_j^{\skpd{m_1-m_0}{n_j}}$.

Consider such a divisor splitting into nontoric divisors as
$$i^\star D_1 = \sum_j \tilde{D}_1^{(j)}.$$
The embedding of each nontoric component into $X_\Delta$ is given
by the intersection of $D_1$ with the set $\{r = z_j\}$, where $r$
is defined as above and $z_j$ is the corresponding zero of the
polynomial $\chi|_{z_i = 0}$ considered as a function of $r$. This
description implies, that
\[\forall j,k: i_\star \tilde{D}_1^{(j)} = i_\star
\tilde{D}_1^{(k)},\label{embeddedequiv}\]
 since the rational functions $r - z_j$
and $r - z_k$ only differ by a constant and the difference of
their sets of zeroes consequently is just the principal divisor
defined by the rational function $\frac{r - z_j}{r - z_k}$.

Using the explicit description it is clear that
$$i^\star D_1 \cdot i^\star D_2 = 0\;\Rightarrow \; \forall j:
\tilde{D}_1^{(j)} \cdot i^\star D_2 = 0.$$
 If $i^\star D_2$ also splits into several nontoric divisors, it
is also clear that
 $$i^\star D_1 \cdot i^\star D_2 = 0\;\Rightarrow \; \forall j,k:
\tilde{D}_1^{(j)} \cdot \tilde{D}_2^{(k)} = 0.$$

As the restricted polynomial is the same for all toric divisors
corresponding to points in the interior of the same edge
$\theta^\ast$, for neighboring points $n_1, n_2$ one has:
\[\tilde{D}_1^{(j)} \cdot \tilde{D}_2^{(k)} = 0 \:\mbox{whenever $j \not=
k$}.\label{internontoriczeros}\]

Since $i_\star$ maps the class of a point to the class of a point,
using (\ref{embeddedequiv}) one can deduce
\begin{eqnarray*}
\forall j,k:\:\tilde{D}_1^{(j)} \cdot i^\star D_2 & = & i_\star
(\tilde{D}_1^{(j)} \cdot i^\star D_2) \\
 & = & D_2 \cdot i_\star \tilde{D}_1^{(j)} \\
 & = & D_2 \cdot i_\star \tilde{D}_1^{(k)} \\
 & = & i_\star (\tilde{D}_1^{(k)} \cdot i^\star D_2) =
 \tilde{D}_1^{(k)} \cdot i^\star D_2\\
\end{eqnarray*}
and therefore
\[\tilde{D}_1^{(j)} \cdot i^\star D_2 = \frac{1}{l(\theta)} i^\star
D_1 \cdot i^\star D_2.\]
 Because of (\ref{internontoriczeros}) we thus obtain
\begin{eqnarray}
\tilde{D}_1^{(j)} \cdot \tilde{D}_1^{(k)} & = &
\delta_{j,k}\,\tilde{D}_1^{(j)} \cdot i^\star D_1 \nonumber \\
& = & \delta_{j,k}\, \frac{1}{l(\theta)} i^\star D_1 \cdot i^\star
D_1
\end{eqnarray}
for the intersection numbers of nontoric divisors belonging to the
same toric divisor. For the intersections of the nontoric
components of divisors belonging to neighboring points $n_1, n_2$
on an edge of $\Delta^\ast$ we finally obtain
\begin{eqnarray}
\tilde{D}_1^{(j)} \cdot \tilde{D}_2^{(k)} & = &
\delta_{j,k}\,\tilde{D}_1^{(j)} \cdot i^\star D_2 \nonumber \\
& = & \delta_{j,k}\, \frac{1}{l(\theta)} i^\star D_1 \cdot i^\star
D_2.
\end{eqnarray}

\section{Main result and computer proof} \label{mainandalgo}

We are now ready to state our main result. First note that by using a
toric hypersurface as string compactification space we imply taking
the metric and $B$ field as the restriction of corresponding data
on the ambient toric variety\footnote{In e.g. Witten's linear
sigma model approach \cite{wittenphases} this restriction is
implicitly contained.}.

\begin{prop} \label{proofk3mirror}
Let $\Delta$ be a reflexive three dimensional polyhedron. The
family of smooth toric K3 hypersurfaces in $X_\Delta$ (together
with complexified K\"ahler form) is an analytically open subset of
the family $K3_{Pic_{tor}(\Delta), Pic(\Delta)}$ as defined in
section \ref{mirsymorthpic}. The following equations hold:
\begin{enumerate}
\item $Pic_{\mathrm{tor}}(\Delta) = (Pic(\Deltastar))^\perp \subset
\tilde{H}^2(X,\ZZ)$ and
\item $Pic_{\mathrm{tor}}(\Deltastar) =
(Pic(\Delta))^\perp \subset \tilde{H}^2(X,\ZZ)$
\end{enumerate}
\end{prop}

The first part of proposition \ref{proofk3mirror} follows from the
definition (c.f. section \ref{mirsymorthpic}) and counting
dimensions. Dimensional counting also shows, that the set of all
twoplanes $\Omega$ corresponding to smooth toric K3 hypersurfaces
in $X_\Delta$ (i.e. obtainable by polynomial deformations) is a
Zariski open subset of the set of twoplanes in
$Pic(\Delta)^\perp$. The restriction to the analytically open
interior of a bounded domain only occurs for the twoplanes $\mho
\subset Pic_{tor}(\Delta) \tensor \RR$. Hence, we obtain

\begin{cor}
Analytically open subsets of the families of smooth K3
hypersurfaces in $X_\Delta$ and $X_{\Delta^\ast}$ are mirror dual.
\end{cor}

\begin{rem} The above restriction to open subsets can be lifted by
consideration of additional phases of the corresponding theories.
These phases are closely correlated with the secondary fans of
$\Delta$ and $\Deltastar$. Since at least some of the occuring
phases cannot be described as nonlinear K3 sigma models, this is
clearly outside the scope of this paper.
\end{rem}

What remains to prove is
 \[Pic_{\mathrm{tor}}(\Delta) =
(Pic(\Deltastar))^\perp \subset
\tilde{H}^2(X,\ZZ).\label{toprove}\]

As $\tilde{H}^2(X,\ZZ) = -E_8 \oplus -E_8 \oplus H \oplus H$ is an
even selfdual lattice just like $H^2(X,\ZZ)$ itself, the following
tools will be useful:

\begin{defn} \label{discrgroupdef}
Let $\Gamma$ be an even, nondegenerate lattice (i.e.\ $\forall x
\in \Gamma:\;\skpd{x}{x} \in 2\ZZ$ and $\skpd{x}{y} = 0\,\forall
y\in\Gamma \Rightarrow x = 0$). One then has a natural embedding
of $\Gamma$ into its dual lattice $\Gamma^\star$ and because of
$\Gamma^\star \subset \Gamma \otimes \QQ$ the scalar product on
$\Gamma$ can be uniquely extended to $\Gamma^\star$. The
\textbf{discriminant group} of $\Gamma$ is defined to be
$$G_D(\Gamma) \defas \Gamma^\star / \Gamma.$$
\end{defn}

Obviously, the discriminant group is finite and abelian. Using the
scalar product on $\Gamma$ one can define a quadratic form on the
discriminant group:

\begin{defn}
Let $\Gamma$ be as in definition \ref{discrgroupdef}. On the
discriminant group $G_D(\Gamma)$ one defines a quadratic form $q:
G_D(\Gamma) \rightarrow \RR\,\mathrm{mod}\,2\ZZ$ as follows: Let
$x\in\Gamma^\star$ be a representative of $\bar{x}\in
G_D(\Gamma)$. We set
$$q(\bar{x}) \defas \skpd{x}{x}\,\mathrm{mod}\,2\ZZ.$$
 $q$ is well defined, because for $x\in \Gamma^\star$ and $y\in\Gamma$
$$\skpd{x+y}{x+y}-\skpd{x}{x} = 2 \skpd{x}{y} + \skpd{y}{y} \in
2\ZZ.$$ $q$ is called the \textbf{discriminant form}.
\end{defn}

\begin{lem} [\cite{nikulinlemma}] \label{nikulindisklemma}
Let $L \subset \Gamma$ be a primitive sublattice of an even,
selfdual lattice $\Gamma$. Let $L^\perp$ denote its orthogonal
complement and assume $L \cut L^\perp = \{0\}$. Then $L^\star / L
\cong (L^\perp)^\star / L^\perp$ and the quadratic forms only
differ by sign. Conversely, if $L, L^\perp$ are nondegenerate even
lattices with the same discriminant form up to sign (we will
denote the isomorphism by $\gamma$), then $L^\perp$ is the
orthogonal complement of $L$ in
$$\Gamma \defas \left\{(l,l') \in  L^\star \oplus
(L^\perp)^\star \mid \gamma(\bar{l}) = \bar{l}' \right\},$$ where
$\bar{l}$ denotes the image of $l$ under the quotient map $L^\star
\rightarrow L^\star / L$.
\end{lem}

Using arguments as in \cite{batdualpoly} one can show that the
objects in (\ref{toprove}) have the right dimensions\footnote{Use
$\rank(i^\star A^1(X_\Delta))+3 = \dim H^{1,1}(X^a_{\Delta^\ast})
= \#\{\mbox{integer points on faces of}\,\Delta\,\mbox{with codim}
\ge 2\}$ and the exact sequence from the proof of theorem 4.4.2.}.
However, this is of course not sufficient to prove
(\ref{toprove}). The formulas given in section \ref{toricalgo} are
sufficient to calculate the lattices in (\ref{toprove}) for any
given reflexive $\Delta$, but unfortunately do not allow for a
general proof.

Here, the known classification of all three dimensional reflexive
polyhedra in \cite{kreuzskarkek3list} comes to the rescue. As
there are only 4319 reflexive polyhedra in three dimensions, one
can show (\ref{toprove}) for each reflexive $\Delta$ seperately.

Although the necessary calculations are simple enough to be done by
hand, the number of polyhedra to check suggest delegating this
work to a computer, in particular since the used classification
has also been done by computer.

For the sake of using lemma \ref{nikulindisklemma}, we can still
prove the following\footnote{This fact can easily be
checked while explicitly calculating the lattices -- nevertheless
a general proof is nicer.}

\begin{lem}
Let $\Delta$ be a reflexive polyhedron, $\dim \Delta = 3$. Then
both $Pic(\Delta)$ and $Pic_{\mathrm{tor}}(\Delta)$ are even
nondegenerate lattices.
\end{lem}
\textbf{Proof:} As sublattices of $H^2(Z(\chi), \ZZ)$ both
lattices are even. $Pic(\Delta)$ is nondegenerate due to
Poincar\'e duality on the K3 surface. Let $Pic_{\mathrm{tor}}(\Delta)
\ni x = i^\star \tilde{x},\,\tilde{x}\in
A^1(X_\Delta).$

Suppose $\forall y\in Pic_{\mathrm{tor}}(\Delta):
x\cdot y = 0\;\Leftrightarrow\;\forall \tilde{y}\in A^1(X_\Delta):
\tilde{x} \cdot Z(\chi) \cdot \tilde{y} = 0$. Due to Poincar\'e
duality on the smooth toric variety $X_\Delta$, $\tilde{x} \cdot
Z(\chi) = 0\;\Rightarrow\;\tilde{x}\in \ker i^\star
\;\Rightarrow\; x=0$.\hfill$\Box$
\par
\hfill\\It now suffices to perform the following calculation for
each $\Delta$ from the classification list:
\begin{enumerate}
\item Calculate $Pic_{\mathrm{tor}}(\Delta)$ and
$Pic(\Deltastar)$ using the formulas in sections
\ref{toricpiccalc} and \ref{completepicard}.
\item
\label{fudamentaltrafo} Calculate the discriminant groups and
forms. In this step one uses finiteness of the discriminant group,
which is generated by the equivalence classes of the generators of
the dual lattice (which we denote by $L^\star$). $L^\star / L$ can
then be constructed by
repeatedly adding the equivalence classes of these generators. To
this end one starts with $0 \in L^\star$, which yields the unit of
the discriminant group. One now constructs a list of all group
elements by adding all generators of $L^\star$ to representatives
of all group elements already contained in the list. The newly
constructed lattice points are then transformed into the
fundamental domain of the sublattice $L \subset L^\star$ and added
to the list, if they are not already contained. Obviously the
algorithm stops when representatives for all elements of the
discriminant group are contained in the list. After this
calculation, the group multiplication table and the quadratic form
are readily calculable.
\item \label{detcyclic} Split the discriminant groups into their cyclic
factors\footnote{This is possible for any finite abelian group.
One possible proof of this well known fact consists of studying the
steps of the algorithm presented in this section.} $\ZZ_{p^q}$
with $p$ prime. To this end one first determines the length of the
orbits of all group elements, thereby splitting the discriminant
group into factors $G_p \defas \bigoplus_k \ZZ_{p^{q_k}}$ for the
pairwise different occurring prime numbers $p$. Beginning with
orbits of maximal length, each $G_p$ is then split into its cyclic
factors $\ZZ_{p^q}$. If the discriminant groups under
consideration split into (modulo reordering) different cyclic
factors, we can stop the calculation as the discriminant groups
are different.
\item As the embedding of the cyclic factors is not uniquely determined,
one now constructs an exhaustive list of isomorphism candidates by
trying to map the generators of the cyclic factors of the first
discriminant group (as chosen during step \ref{detcyclic} above)
to elements of the second one with equal orbit length, thereby
defining isomorphisms of cylclic subgroups. If the second
discriminant form evaluated on the images differs from the first
one evaluated on the preimages only by sign, one finally checks
the constructed collection of isomomorphisms of cyclic subgroups
for being an isomorphism of the discriminant groups by
constructing the whole image and comparing multiplication
tables\footnote{The order of the tests is of course not mandatory
-- it just makes the algorithm faster to perform the easiest
checks first.}. If an isomorphism preserving the discriminant form
(up to sign) is found, we are done. If all possible mappings are
checked and such an isomorphism is not found, (\ref{toprove}) must
be false.
\par For cases with many isomorphic cyclic factors this step takes the
most computing time.
\end{enumerate}

The described algorithm was implemented using C++ and (using an
800 MHz PC) applied to all 4319 reflexive polyhedra from the
classification list. After roughly two hours of computing time all
discriminant groups were calculated and for all cases suitable
isomorphisms were successfully determined. This proves proposition
\ref{proofk3mirror} by explicit calculation.

A list of the found discriminant groups and forms can be found at
\cite{frk3disclist}.

\begin{rem}
Specifying the matrix of the discriminant form on the generators
of the discriminant group completely determines the full
discriminant form, because evaluating the scalar product on
representatives of the group elements is well defined modulo $\ZZ$
(the offdiagonal elements of the matrices in the above list are
therefore only well defined modulo $\ZZ$, not modulo $2\ZZ$!).
This nevertheless allows calculating the squares of all other
group elements:

If we choose a basis for the even nondegenerate lattice $L$ and
the corresponding dual basis for $L^\star$ and denote the matrix
of the scalar product on $L$ by $g$, the embedding $L
\hookrightarrow L^\star$ is given by the matrix $g$ and the
induced scalar product on $L^\star$ is given by $g^{-1}$. We write
$\{l_k\}$ for an arbitrarily chosen collection of representatives
of generators of $L^\star / L$.

Modulo $2\ZZ$ we then have:
$$\forall a_i \in \ZZ: g^{-1}(\sum_k a_k l_k, \sum_k a_k l_k) =
\sum_k a_k^2 g^{-1}(l_k,l_k) + \sum_{j\not=k} 2 a_j a_k
g^{-1}(l_j, l_k).$$ Hence, the scalar products between the
generators only have to be defined modulo $\ZZ$ in order to
reconstruct the complete quadratic form. The latter is true
because modulo $2\ZZ$ $\forall l' \in L$:
$$g^{-1}(l_j + g l', l_k) = g^{-1}(l_j, l_k) + g^{-1}(g l', l_k)$$
and $g^{-1}(g l', l_k) \in \ZZ$.
\end{rem}

\begin{rem}
The transformation into the fundamental domain in step
\ref{fudamentaltrafo} of the algorithm can be done as follows:

Let a matrix $\Lambda$ define an embedding of some sublattice of
rank $d$ into $\ZZ^d$ (here $\Lambda = g$). A lattice point $x \in
\ZZ^d$ lies in the fundamental domain given by $\Lambda$, if and
only if (over $\QQ$)
$$\Lambda^{-1} x \in [0,1)^d.$$
Hence, transforming any lattice point $x$ into the fundamental
domain can be done by setting
$$\bar x \defas \Lambda(\Lambda^{-1}x \,\mathrm{mod}\, \ZZ) =
x - \Lambda \lfloor \Lambda^{-1}x \rfloor \in \ZZ^d.$$
 This calculation can be done using only integers, if one
calculates the inverse of $\Lambda$ (e.g. using Gaussian
elimination) over $\ZZ$, i.e. $\Lambda^{-1} =
\frac{\tilde{\Lambda}}{\lambda}$, $\tilde\Lambda$ and $\lambda$
integer. Then
$$\bar{x} = \Lambda(\frac{\tilde{\Lambda}}{\lambda} x\,\mathrm{mod}\, \ZZ)
= \frac{1}{\lambda} \Lambda(\tilde{\Lambda} x\,\mathrm{mod}\,
\lambda \ZZ).$$
\end{rem}

\begin{rem}
Some nice explicit examples calculated using the methods in a
preliminary version of this paper can be found (embedded into a
larger context) in \cite{katrinlink}.
\end{rem}

\section*{Acknowledgements}
The author wants to thank V.V. Batyrev, W. Nahm, K. Wendland and
D. Roggenkamp for helpful discussions and comments.

\newcommand{\etalchar}[1]{$^{#1}$}

\end{document}